\documentclass[prd,aps,twocolumn,preprintnumbers, showpacs, nofootinbib,superscriptaddress,notitlepage]{revtex4-1}
\usepackage{mathrsfs}
\usepackage{amsfonts}
\usepackage{amsmath}
\usepackage{slashed}
\usepackage{array}
\usepackage{verbatim}
\usepackage{epsfig}
\usepackage{graphicx}
\usepackage{color}
\usepackage[dvipsnames]{xcolor}
\usepackage{caption}
\usepackage{subcaption}

\newcommand{\beq}{\begin{eqnarray}}
\newcommand{\eeq}{\end{eqnarray}}

\def\be{\begin{equation}}
\def\ee{\end{equation}}
\def\bea{\begin{eqnarray}}
\def\eea{\end{eqnarray}}

\begin{document}

\title{UV divergence of the quasi-PDF operator under the lattice regularization}

\author{Yi-Kai Huo}
\affiliation{Zhiyuan College, Shanghai Jiao Tong University, Shanghai 200240, China}

\author{Peng Sun}
\affiliation{Nanjing Normal University, Nanjing, Jiangsu, 210023, China}

\date{\today}

\begin{abstract}
Even since the ``quasi" parton distribution function (PDF) was proposed under the large-momentum effective theory (LaMET) framework, its renormalization under the lattice regularization has been a central challenge to be solved due to the linear divergence. Thus, we investigate several possible ways to renormalize the quasi-PDF operators in high accuracy with non-perturbative calculation using the quench configurations at several lattice spacings. We find that the ratio of the UV divergences obtained from the Wilson loop and off-shell quasi-PDF operator is not a constant of the Wilson link length $z$. Although the linear divergence in them may be consistent to each other numerically, there is some additional UV divergence in the quasi-PDF operator.
\end{abstract}

\maketitle

\section{Introduction}
Parton distribution function (PDF) is an effective way to depict the probability density of quark and gluon inside a high-speed nucleon.
Such a function plays an essential role in the long-distance part of the hadron collision cross sections, and its uncertainty has been one of the major
sources of the error in the theoretical predictions of cross sections at LHC or other hadron colliders. Thus it is important for the accurate test of Standard Model(SM) in the search of the tiny beyond SM signal. On the other hand, it directly relates to the density of different partons in the nucleon, and provides the direct information on the intrinsic properties of nucleon like the origin of the proton spin and whether the sea quarks play important roles. Beyond PDF, there are also kinds of the functions are proposed to investigate the 3D pictures of nucleon, such as generalized parton distributions, transverse momentum dependent distributions and so on.

Until recently, the only reliable way to obtain PDF is the global fit of experimental data based on certain parameterized form, as the typical scale of nucleon is  $\Lambda_{QCD}$ and then the perturbative calculation is impossible. On the other hand, PDF cannot be defined in the Euclidean space lattice and then cannot be directly accessed by the non-perturbative calculation likes Lattice QCD. Thus, the large-momentum effective theory (LaMET) suggests a possible way to calculate the quasi-PDF operators
\bea\label{eq:quasi}
\tilde{h}(z, P_z, \tilde{\mu}) = \langle P|\bar{\psi}(z) \gamma_t W(z, 0) \psi (z) |P \rangle
\eea
 on the lattice, where $|P\rangle$ is the nucleon state with momentum $P$, and $W(z, 0) = \exp(-ig\int_0^{z} dz' A_z(z'))$ is the Wilson link along the $z$ direction. And then we can extract PDF through the perturbative matching between the Fourier transform of the renormalized quasi-PDF and PDF~\cite{Ji:2013dva}.

But unlike the dimensional regularization widely used in the continuum calculation, the quasi-PDF under the lattice regularization can have linear divergence, and then such divergence should be removed properly before we match the lattice calculation to PDF~\cite{Xiong:2013bka}.
The theoretical studies so far show that the linear divergence only comes from the Wilson link $W$~\cite{Ji:2015jwa, Ji:2017oey,Ishikawa:2017faj,Green:2017xeu} and the quasi-PDF operator is multiplicative renormalizable, thus we can simply extract the linear divergence from the Wilson loop and use it into the renormalization of the quasi-PDF operator~\cite{Chen:2016fxx,Zhang:2017bzy}, or obtain the entire UV divergence of the quasi-PDF operator through its matrix element in the off-shell quark state using the regularization independent momentum subtracted (RI/MOM) schemes ~\cite{Green:2017xeu,Chen:2017mzz,Alexandrou:2017huk}.
In principle, both of the methods will give us the same results after they are extrapolated to the continuum limit; but an accurate numerical test is still absent so far.  In this work, we carry out the calculation of both methods with the lattice spacing from $0.16$ fm to $0.08$ fm using quench configurations, and investigate their differences.

The rest of this paper is organized as follows. In
Sec.~\ref{sec:prm}, we describe the procedure of the UV divergence calculation on the quasi-PDF through Wilson loop and also off-shell quark matrix element with lattice QCD; then the numerical details of the calculation and comparison are provided in Sec.~\ref{sec:calc}. A brief summary is presented in the last section.

\section{Numerical details}
\label{sec:prm}


Since a quark propagator is equivalent to a Wilson line in the heavy-quark limit, and then a Wilson loop $W(r, t)$ can be considered as a heavy quark and antiquark pair where $r$ and $t$ are the lengths in space and virtual time~\cite{Chen:2016fxx}. Thus, the statical potential can be extracted from the decay of the Wilson loop with time t,
\bea\label{eq:potential1}
V(r)=-\frac{1}{a}\lim_{t\rightarrow \infty}\ln\frac{\mathrm{Tr}[\langle W(r,t)\rangle]}{\mathrm{Tr}[\langle W(r,t-a)\rangle]}\label{eq:vr},
\eea
where $a$ is the lattice spacing. It can be parameterized into the Cornell potential form
\bea\label{eq:potential2}
V(r) =- \frac{e_c}{r}+\frac{2\delta m}{a}  + \sigma  r +V_{0},
~\label{eq:staticp}
\eea
and then four terms in the right hand side correspond to the Coulomb term with the coupling $e_c$, linear divergence term with the coefficient $\delta m$ (factor 2 for the 2 more units of the Wilson link in the numerator of the right hand side comparing to the dominator), linear potential term with the string tension $ \sigma$, and the possible residual constant. Those parameters can be obtained from a correlated minimal-$\chi^2$ fit to $V(r)$.

It is obvious that $W(r,t)$ with on-axis $r$ can be obtained by
\begin{eqnarray}
  &W(r,t)&=\langle\sum_x\textrm{Tr}\big[\prod_{i=1}^rU(x+(i-1)\hat{n}_z,x+i\hat{n}_z)\nonumber\\
&&\prod_{i=1}^tU(x+r\hat{n}_z+(i-1)\hat{n}_t,x+r\hat{n}_z+i\hat{n}_t)\nonumber\\
&&\prod_{i=1}^rU^{\dagger}(x+(r-i)\hat{n}_z+t\hat{n}_t,x+(r-i+1)\hat{n}_z+t\hat{n}_t)\nonumber\\
&&\prod_{i=1}^tU^{\dagger}(x+(t-i)\hat{n}_t,x+(t-i+1)\hat{n}_t)\big]\rangle.
\end{eqnarray}
We also calculate the Wilson loop with the off-axis $r$ by taxi-driver links, e.g., the Wilson link with $r=\sqrt{2}$ can be obtained by
\bea
&U(x,x+\hat{n}_x+\hat{n}_y)&\equiv U(x,x+\hat{n}_x)\nonumber\\&&U(x+\hat{n}_x,x+\hat{n}_x+\hat{n}_y).
\eea
Since the relative uncertainty of $W(r,t)$ will exponentially increase with $t$,  we fit the $W(r,t)$ using the empirical form
\bea
W(r,t) =& C_{0} e^{-V(r)t}(1 + C_1 e^{-\Delta V t})
\eea
 to get $V(r)$ with reasonable statistical uncertainty.

For the Quasi-PDF operator
\bea
{\cal O}(z)\equiv\bar{\psi}(z) \gamma_t W(z, 0) \psi (z)
\eea
, it claimed that the power divergence in the Wilson link can be canceled by a factor $\mathrm{e}^{-\delta m z/a}$~\cite{Ji:2017oey,Ishikawa:2017faj}, where $z$ is the link length, a is the lattice spacing and $\delta m$ is exactly what defined in Eq.~(\ref{eq:potential2}). At the same time, the RI/MOM renormalization constant of ${\cal O}(z)$ can be obtained by requiring loop corrections for the matrix element of a quasi-PDF operator vanish in an off-shell quark state at a given momentum:
\begin{align}\label{eq:Z}
Z(z,a^{-1},\mu_R)=\left.\frac{\sum_s\langle p,s|O_{\gamma^t}(z)|p,s\rangle}{\sum_s\langle p,s|O_{\gamma^t}(z)|p,s\rangle_{\rm tree}}\right|_{\tiny\begin{matrix}p^2=-\mu_R^2 \\ \!\!\!\!p_z=0\end{matrix}}\,.
\end{align}
The bare matrix element $\sum_s\langle p,s|O_{\gamma^t}(z)|p,s\rangle$ will be calculated on the Lattice from the amputated Green's function $\Lambda_{\emph{O}}(p,z)$ with proper projection,
\bea
\Lambda_{\emph{O}}(p,z) &=& S^{-1}(p)\langle \sum_{w} \gamma^5 S^{\dag}(p, w+z\hat{n_z}) \nonumber\\
&&\ \ \ \ \gamma^5 \Gamma W(w+z\hat{n},w)S^{\dag}(p, w)\rangle S^{-1}(p)
\eea
where $\hat{n_z}$ is a unit vector along z-axis, the middle part will be calculated over the whole lattice space and finally summed up. As for the quark propagators, they are defined by
\be
\begin{aligned}
S(p, x)= \sum_y \langle \bar{\psi}(y) \psi(x) \rangle,\
S(p)= \sum_x S(p,x)
\end{aligned}
\ee

Eventually we will consider the ratio
\bea\label{eq:ratio}
R(z,a)=e^{\delta m z/a} Z(z,a^{-1},\mu_R),
\eea
which should be insensitive to $a$ up to the $\textrm{log}(a)$ and $a^2$ corrections, if all the linear divergences are cancelled.

\section{Lattice results}\label{sec:calc}

  \begin{figure}[tbp]
  \centering
  \includegraphics[width=8cm]{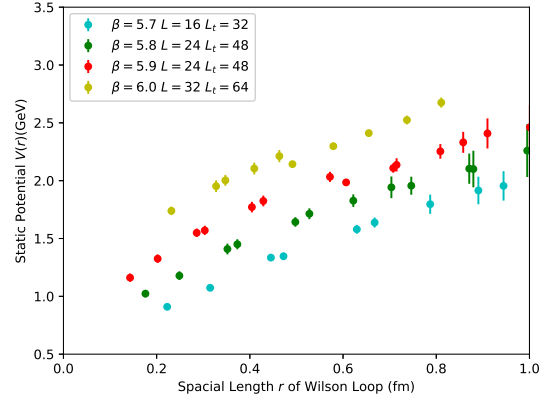}
  \caption{Static potential variation trend of spacial length on different lattices}
  \label{fig:pot}
  \end{figure}

We generate four quenched ensembles with the CHROMA package using different values of $\beta=6/g_0^2$, 5.7, 5.8, 5.9 and 6.0. In order to reduce the effect of the correlation between the neighboring configurations, we save one configuration per 5 sweeps after a 300-sweep warm up process. 100 configurations are generated for each ensemble. The Wilson loops $W(r,t)$ are calculated with $r=1,2,\ldots, 10$, $\sqrt{2}, 2\sqrt{2}, \ldots, 7\sqrt{2}$, and also $\sqrt{3}, 2\sqrt{3}, \ldots, 5\sqrt{3}$, and fit the $V(r)$ using the range $t\in[1,10]$. In the calculation of $W(r, t)$, we applied the APE smearing in the spacial direction while keep the time direction untouched. The lattice spacing is obtained by
\begin{equation}
a\equiv 0.46 \sqrt{\frac{\sigma a^2}{1.65-e_c}}\mathrm{fm}
\end{equation}
based on the fit of $V(r)$ using the functional form defined in Eq~(\ref{eq:potential2}). The fitting result of $\delta m$ is
\bea
\delta m = 0.3405 \pm 0.0005 \mathrm{fm}^{-1}
\eea
Using those lattice spacings, we tune the quark mass parameter $m_{\mathrm{q}}$ and clover parameter $C_{\mathrm{SW}}$~\cite{Gupta:1997nd}, to make the pion mass used by the quasi-PDF calculation at all the ensemble to be roughly 500 MeV. The setup of simulation are collected in Table~\ref{tab:lattice} and Fig.~\ref{fig:pot} shows the $V(r)$ we obtained at different lattice spacings and the values are rescaled to the physical units.

\begin{table}[htbp]
  \centering
  \begin{tabular}{ccccccc}
  \toprule
  $\beta$ & $L$ & $L_t$ & $C_{\mathrm{SW}}$ & $m_{q}$ & $a_{\mathrm{phy}}(\mathrm{fm})$\\
\hline
  5.7 & 16 & 32 & 1.5675313233   & -0.473 & 0.1574 \\
\hline
  5.8 & 24 & 48 & 1.52907964207  & -0.437 & 0.1244  \\
\hline
  5.9 & 24 & 48 & 1.50105351132  & -0.413 & 0.1011 \\
\hline
  6.0 & 32 & 64 & 1.47851437189  & -0.395 & 0.0819  \\
\hline
  \end{tabular}
  \caption{Quenched ensemble setup}
  \label{tab:lattice}
\end{table}

\begin{figure}[tbp]
  \centering
  \includegraphics[width=8cm]{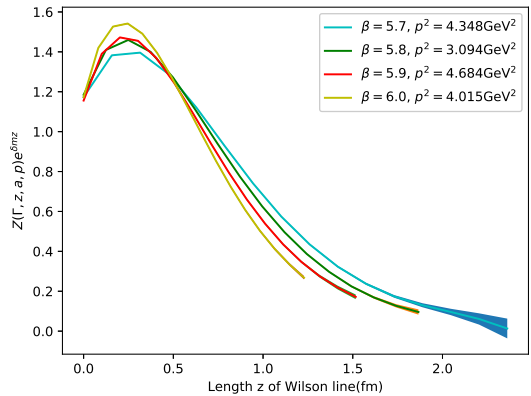}
  \caption{$Z(\Gamma, z, a, p)e^{\delta mz}$ of individual $z$ where the red line stands for the mean value of the ratio, the blue band shows the range of uncertainty}
  \label{fig:ratio}
  \end{figure}

In principle, we should keep $p^2$ to be the same on all the ensembles to preform a fair comparison. But it is not doable as the physical volume of the ensembles we have are not exactly the same. Thus we just set a momentum parameter $p=(3,3,0,0)2\pi/L$ for all sets of configurations, as the $p^2$ dependence is weak. The ratio $R$ defined in Eq.~(\ref{eq:ratio}) is plotted in Fig.~\ref{fig:ratio}. It is obvious that there is some residual effect at large $z$ after the linear divergence is removed.

\section{Summary}

We compare the renormalization of Quasi-PDF though two popular ways, one is by extracting the power divergence coefficient $\delta m$ though Wilson loops, the other one is by calculating NPR constant in RI/MOM scheme. In order to investigate the difference between these two methods, we generate configurations on different quenched lattices.
In order to investigate the difference between both methods, we
generate configurations on the quenched lattices, and then extract the UV divergence pieces through these two methods.
We find that the ratio of the two UV divergence pieces is not a constant naively but proportional to the length of the Wilson
line. It means the UV behavior in Wilson loop and  Quasi-pdf are not exactly same. Although the
linear divergence in them may be consistent to each other numerically, there is some  additional UV divergence
in bare quasi-pdf.

\section*{Acknowledgement}
The authors thank Yi-Bo Yang for valuable comments. The LQCD calculations were performed using the Chroma software suite~\cite{Edwards:2004sx}.
The numerical calculation is supported by Jiangsu Key Lab for NSLSCS and Center for HPC of Shanghai Jiao Tong University.
P. Sun is supported by Natural Science Foundation of China under
grant No. 11975127.
%
\end{document}